\documentstyle[graphicx,preprint,prb,aps]{revtex}

\begin{document}
\draft
\title{
From nonwetting to prewetting: the asymptotic behavior of 
$^4$He drops on alkali substrates
}
\author{M. Barranco,$^1$ M. Guilleumas,$^1$ E.S. Hern\'andez,$^2$
R. Mayol,$^1$ M. Pi,$^1$ and L. Szybisz$^{2,3}$}
\address{$^1$Departament ECM,
Facultat de F\'{\i}sica, Universitat de Barcelona, E-08028 Barcelona,
Spain
}
\address{$^2$Departamento de F\'{\i}sica, Facultad de Ciencias Exactas y Naturales,
Universidad de Buenos Aires, 1428 Buenos Aires, and Consejo Nacional de
Investigaciones Cient\'{\i}ficas y T\'ecnicas, Argentina}
\address{$^3$ Departamento de F\'{\i}sica, CAC,  
Comisi\'on Nacional de Energ\'{\i}a At\'omica, 1429 Buenos Aires, Argentina}
\date{\today}

\maketitle

\begin{abstract}

We investigate the spreading of  $^4$He 
 droplets on alkali surfaces at zero temperature, within the frame 
of Finite Range Density Functional theory. 
The equilibrium configurations of  several $^4$He$_N$
 clusters and their asymptotic trend with 
increasing particle number $N$, which can be traced to the wetting behavior of
 the quantum fluid, are examined for nanoscopic 
droplets. We discuss the size effects, inferring that the asymptotic
 properties of large droplets correspond to those of the prewetting film.

\end{abstract}


\section{Introduction}

Wetting phenomena\cite{degennes85} of alkali and graphite-based surfaces
 by quantum fluids have been a burgeoning field of research in recent years, 
both from the experimental and the theoretical viewpoints.  
Early in the last decade, it was  demonstrated that liquid  $^4$He is not  a 
universal wetting agent,
\cite{nacher91,taborek92,bigelow92,mukherjee92,ketola92,rutledge92,taborek94}
  since it fails to spread uniformly  on cesium
surfaces at temperatures below 1.95 K. 
 From the theoretical viewpoint,  Cheng {\it et al}\,\cite{cheng91} predicted,
in the frame of a  nonlocal density functional model, the inability of
helium    to   wet   all heavy   alkalis.   More    stringent   theoretical
studies\cite{cheng92a,cheng92b,saam92,cheng93b,clements94} showed that
while for the majority of moderately thin to thick   helium films
 on substrates, one or two
layers  of solid  helium would  be continuously  wetted by  the excess
liquid, a different pattern was to be expected in the presence of weak
adsorbers. In this case, either nonwetting,  or wetting  preceded by  a first
order prewetting  transition,\cite{cahn77}    
 was most  likely to
appear. The prewetting transition undergone by $^4$He on Cs substrates
was  first observed  by  Hallock and  coworkers\cite{ketola92} and  the
complete  phase diagram  was presented  in
 Ref.\onlinecite{rutledge92}. The situation is not so clear for
 Rb surfaces, since although  a number of experimental 
evidences\cite{bigelow92,mistura94} and at least one
theoretical  calculation\cite{szybisz00b} 
 are consistent with a  vanishing wetting temperature,  
 a  more recent
 experiment\cite{klier02} indicates that pristine Rb is nonwetted by
 helium up to a temperature above 300 mK, in agreement with earlier 
 predictions.\cite{chimezhya98,ancilotto00} 

So far, theoretical work on wetting properties of helium
relies almost exclusively on the description of structure and energetics of
films, uniformly extended on the plane perpendicular to the substrate, in spite
 that all these systems, even on wetted substrates, are thermodynamically 
unstable for the lowest areal coverages and should then appear as
collections of
 puddles or clusters upon the adsorbing surface.  In a prior  work, 
Ancilotto {\it et al}\,
\cite{ancilotto98} have presented  a calculation  of density  profiles of  finite
droplets of $^4$He  on  Cs, at zero temperature.  These authors  solve  a  
nonlinear  equation for  the
density profile, constructed by functional differentiation of a Finite
Range Density Functional  (FRDF) acknowledged as
the Orsay-Trento  (OT) density  functional.\cite{dalfovo95} Quite elaborated 
 from the numerical  viewpoint,  this  work  develops  an  accurate  
theoretical instrument to  investigate wetting  patterns.

The purpose of this work is to perform a  detailed
investigation of the spreading of $^4$He  
droplets on alkali surfaces at zero temperature
in the framework of the FRDF formalism. Our systematic study
permits us to relate the asymptotic
trends of their energetics and structure, with increasing size, 
to the wetting properties of the fluid; in particular, we show that
droplets
as large as a few thousand $^4$He atoms, on the strongest adsorbers,
already
exhibit the structural and thermodynamic characteristics of the
prewetting film at zero temperature. 

In Sec. II we shortly review the current FRDF formulation and give 
essential details of the method applied to solving  the associated 
Euler-Lagrange (EL) equation to obtain the equilibrium configuration. In
Sec. III we 
illustrate  the predicted prewetting
transition for films on alkali adsorbers, and subsequently, we present
the numerical results for clusters.
Our conclusions and perspectives are presented in Sec. IV.

\section{Density functional description of $^4$He clusters on
adsorbing substrates}

The FRDF for $^4$He adopted in this work is that of
Ref. \onlinecite{mayol01}, which was originally developed to describe mixtures
of helium isotopes,  here restricted to the $^4$He component.
Its detailed form and values of the coefficients entering its definition
have been given in Ref.\onlinecite{barranco97}, with the two 
changes reported in Ref. \onlinecite{mayol01} which consist, on the
one hand, in  the neglect of
 the nonlocal gradient correction to the kinetic energy
term  in Ref. \onlinecite{dalfovo95}, and on the other, in the choice
 of the suppressed Lennard-Jones core as
 in Ref. \onlinecite{dupontroc90}, namely
\begin{equation}
V_{LJ}(r) =\left\{
\begin{array}{lr}
4 \varepsilon  \left[ \left( \sigma_4/r \right)^{12} -
\left(\sigma_4/r\right)^6 \right] &
\quad\rm{if} \ \ r \geq h \\
V_0 r^4  &
\quad\rm{if} \ \ r \leq h \ , \end{array}
\right.
\label{eq1}
\end{equation}
with $\varepsilon=10.22 \,{\rm K}$, $\sigma_4=2.556 \,{\rm \AA}$,
and with hard-core radius $h$= 2.359965 $\rm \AA$. Notice that 
$V_0$ is the value of the 6-12 potential at $r=h$. The value of $h$
has been fixed so that the volume integral of the interaction
$V_{LJ}$ coincides with the one in Refs. \onlinecite{dalfovo95,barranco97}.
This form of the FRDF - we especially comment that dropping  the
nonlocal gradient correction present in the OT density
 functional\cite{dalfovo95} 
increases by around a  3$\%$
the surface tension of $^4$He
with respect to the value there reported-
 has been introduced to render the numerical calculations
less cumbersome. 

The integrodifferential EL equation arising from functional differentiation
of the FRDF is 
\begin{equation}
\left[-\frac{\hbar^2}{2 m_4}\,\nabla^2  +
V\left(\rho\right)\right]\,\Psi({\bf r})
\equiv {\cal H}_H \Psi({\bf r}) =\mu\,\Psi({\bf r})
\label{eq2}
\end{equation}
with $\mu$ the chemical potential that enforces
particle number $N$
conservation in the  drop, and with $\Psi=\sqrt{\rho}$,
where $\rho({\bf r})$ is  the particle density.
The mean field $V({\bf r})$ in the
Hartree Hamiltonian ${\cal H}_H$ includes
the substrate potential $V_s(z)$ chosen to be the appropriate 
Chizmeshya-Cole-Zaremba (CCZ) potential. \cite{chimezhya98}
To illustrate the adsorbing power
of the alkali surfaces, in Table \ref{Tab1} we show the energies
${\cal E}_4$ of one $^4$He atom in the substrate,
 obtained with the CCZ potentials.

In the current geometry, for axially symmetric droplets, we have 
 $\rho({\bf r})=\rho(r, z)$. The EL equation is discretized using
 7-point formulae and solved on a two
 dimensional $(r,z)$ mesh.
 We have used a sufficiently large box with
spatial steps $\Delta r = \Delta z \equiv \Delta = h/12 \sim 0.197 
{\rm \AA}$. The stability
of our results against the increase of the number of mesh points,
as well as with respect to the order of the formulae employed  to discretize
the partial derivatives,  has  been checked comparing the solutions
for spherical drops with those obtained with a spherically symmetric code.
 We have employed an imaginary time method to find
the forward solution of the imaginary time diffusion
equation\cite{Pre92}

\begin{equation}
\frac{\partial \Psi}{\partial \tau} = - ({\cal H}_H - \mu)\Psi
\label{eq3}
\end{equation}
in other words,
\begin{equation}
 \Psi(\tau+\delta\tau) - \Psi(\tau) \equiv \Delta \Psi(\tau)
= - \delta\tau ({\cal H}_H-\mu) \Psi(\tau)  \;\;\; ,
\label{eq4}
\end{equation}
where $\mu=\langle\Psi(\tau)|{\cal H}_H|\Psi(\tau)\rangle$.

To accelerate the self-consistent solution of this
equation, we use the preconditioning smoothing operation
described in Ref.\onlinecite{Hos95}. This means that
$\Delta \Psi(\tau)$ has been smoothed as
\begin{eqnarray}
& &\overline{\Delta\Psi}(r,z;\tau) = \frac{1}{2}\{
\Delta\Psi(r,z;\tau) +
\nonumber
\\
& &
\label{eq5}
\\
& &  \frac{1}{4} [
\Delta\Psi(r-\Delta r,z;\tau) + \Delta\Psi(r+\Delta r,z;\tau) +
\Delta\Psi(r, z-\Delta z;\tau) + \Delta\Psi(r, z+\Delta z;\tau)] \}
\;\;\;   .
\nonumber
\end{eqnarray}

Finally, the perfomance of the code has been further improved by adding a
`viscosity term', i.e., Eq. (\ref{eq4}) has been changed into
\begin{equation}
\Psi(\tau + \delta\tau) = \Psi(\tau) +
\overline{\Delta \Psi(\tau)} +\alpha_V 
[\Psi(\tau) - \Psi(\tau - \delta \tau)]
\,\,\, .
\label{eq6}
\end{equation}
The heuristic viscosity parameter $\alpha_V$ is fixed to a value
of 0.8.

We recall that as shown in Ref. \onlinecite{Pre92}, the maximum 
$\tau$-step, $\delta\tau_m$, that
can be used to produce a stable imaginary time evolution
is $\hbar^2 \delta\tau_m/(2 m_4 \Delta^2) \lesssim
1/4$. The combination of smoothing and viscosity allows one to use
large values of $\delta\tau$, typically  up to
$\delta\tau \sim 0.5 \,\delta\tau_m$. After every $\tau$-step, the
$^4$He density is normalized to $N$.
The iteration procedure starts on the halved density 
of a $^4$He$_{2N}$ cluster calculated with the spherically symmetric
code. Most of the computational time is spent in the
evaluation of the mean field $V(r,z)$ by
folding the helium density with the screened Lennard-Jones
potential. For this reason, the mean field is updated
only every ten $\tau$-iterations.

\section{Equilibrium configurations}

\subsection{Films}

The  thermodynamic criterium for  wetting requests that the surface 
grandpotential, namely at zero temperature
$\sigma= (E - \mu N)/A$, where $A$ is the area of the surface, 
regarded as a function of
 coverage $n = N/A$, displays its absolute minimum when $n$ approaches 
infinity.\cite{cheng92a,cheng93b} For films of $^4$He, and within the density 
functional frame,\cite{szybisz00b} this criterion can be 
formulated in terms of the expansion coefficients of the total energy of the 
film in powers of $1/n$. In particular, it has been shown that the OT density
functional plus the CCZ potential yields, for $^4$He films on alkali 
substrates, wetting of Rb\cite{szybisz00b} and K,\cite{szybisz00c} and
nonwetting of a Cs surface.\cite{szybiszp} 
In Figs. \ref{fig1} and \ref{fig2}, we
respectively plot  the chemical potential and the surface
grandpotential of $^4$He on different alkali surfaces, as predicted by
the current FRDF. The energy per particle $e = E/N$ that corresponds
to the equilibrium density for a given coverage
is also shown in Fig. \ref{fig1} in dashed lines for Cs, Na and Li.

It is worthwhile to recall here that from the thermodynamic relations
\begin{equation}
\mu = e + n \frac{\displaystyle \partial e}{\displaystyle \partial n}
\label{eq7}
\end{equation}
\begin{equation}
\sigma = -n^2\,\frac{\displaystyle \partial e}{\displaystyle
  \partial n}
= \int_0^n d n'\,[\mu(n')-\mu(n)]
\label{eq8}
\end{equation}
one can establish the condition $\mu = e$ to localize the 
prewetting jump. This condition corresponds to vanishing areal
grandpotential
\begin{equation}
\frac{\displaystyle \partial e}{\displaystyle \partial n} = \sigma = 0
\label{eq9}
\end{equation}  
from where the Maxwell construction (\ref{eq8}) defines the
chemical potential and coverage at the prewetting first order transition. 
The prewetting jumps are indicated in Fig. \ref{fig1} as horizontal segments,
 for the stronger adsorbers Na and Li. Although K is wetted by
helium in the current FRDF description (see Fig. \ref{fig3}),
 the prewetting jump lies too
far to the right in the scale here displayed. In the present
calculation, $^4$He does not wet Rb and behaves very
much like Cs; this can be appreciated in Fig. \ref{fig3} where
the surface grandpotential of the film is plotted as a function of
the inverse coverage. Indeed, it has been already pointed out that
this substrate
 represents a delicate limiting case, largely sensitive to the details of the 
calculation, of the density functional and of the adsorbing potential 
adopted.\cite{szybisz00b,ancilotto00,szybiszp}

\subsection{Clusters}

We concentrate on droplet sizes below $N$ = 3000, which illustrate the 
spreading trend on  adsorbers of different strengths and provide a
good representation of the asymptotic behavior as shown below. 
Addressing larger drops requires larger mesh sizes, which is possible
at the obvious price of making the calculations more cumbersome and
time consuming. In  Fig. \ref{fig4} we show the contour plots of the
particle densities in the $(x, z)$ plane for a cluster with
$N$ = 1000, on substrates Cs,  K, Na and Li
-Rb looks very much as Cs, and for this reason we do not
discuss it here-
 (hereafter, lenghts are given in ${\rm \AA}$).
The profiles $\rho(r, z_{min})$ of these drops, at the minimum 
$z_{min}$ of the CCZ potential,
 and  $\rho(0,z)$ along the symmetry axis  are plotted in
Fig. \ref{fig5} for the same  substrates. As we see,
increasing attractiveness provokes a change of shape in two
complementary
 manners, which becomes most noticeable for the clusters on Na and Li, i.e., 
 flattening along the vertical coordinate, and sizeable radial spreading.
 We also visualize a smoothing of the structure, since the 
oscillations in $\rho(0, z)$ evolve from practically three  peaks on Cs and 
K, and two on Na,  to just one
on Li. This feature  does reflect the wetting behavior obtained for $^4$He
on these alkalis, namely nonwetting for Cs, 
and prewetting with a  jump of above three layers on  K, two on Na and one
on Li, 
respectively. In fact,
submonolayer wetting occurs for the strongest adsorber Li.
\cite{cheng93b,clements94,boninsegni99}  

The energetics  as a function of cluster size is illustrated in Fig.
 \ref{fig6}, where we  plot the chemical
potential of the  $^4$He atoms,
and the grandpotential per particle
$ \omega = E / N - \mu$ as functions of $N$ for the above
alkalis. From this figure, we realize that for
wetted adsorbers, there is a tendency -very clear for the strongest substrates 
Na and Li- to saturate $\mu$ at a finite value 
below the bulk value -7.15 K, and $\omega$ at zero value.  In the case of Cs,
the limiting chemical potential points towards the bulk figure, in
 agreement with the nonwetting behavior of $^4$He; notice that
 although $^4$He  wets K, as seen in Fig. 3, within the current scale
 we do not reach values of $\mu$ lower than -7.15 K, since this
 crossing takes place at a much larger number of atoms, likely several
 tens of thousands. 

We now select Na as a test case to examine
the evolution of the density profiles with particle number; as a reference,
 in Fig. \ref{fig7} we depict the contour plots of the particle densities
in the $(x,z)$ plane for several values of $N$  varying between 100
and 3000, where we
already observe a rather flat profile with two ridges, corresponding
 to density oscillations parallel to the substrate. Figure \ref{fig8}
 displays  $\rho(r, z_{min})$ and $\rho(0, z)$ for the above particle numbers.
We appreciate a
definite inclination to  saturate the vertical density profiles with
 two  shells and a limiting height, as well as a tendency  
to flatten the radial dependence into a wedge-like shape, splashing outwards
any extra material. These effects are more pronounced in the case Li,
as can be seen
from Fig. \ref{fig9} where the submonolayer profiles become
manifest; nevertheless, it should be kept in mind that as mentioned in
Ref. \onlinecite{boninsegni99}, DF theory predicts a prewetting jump
at a larger coverage than i.e., path-integral Monte Carlo calculations.

These patterns suggest the existence of an asymptotic coverage at the center of
 large clusters; in fact, as we  define
\begin{equation}
n(r) =2\,\pi\,\int dz\,\rho(r, z)
\label{eq10}
\end{equation}
and plot $n(0)$ as a function of $N$ for all 
substrates under consideration as shown in Fig. \ref{fig10}, the shapes
 of these curves indicate that for wetted adsorbers,
 such an asymptotic value exists. Our data indicate that these
 limiting numbers are ($\mu_{Na}, n_{Na}(0)$) = 
(-8.27 K, 0.14 $\rm \AA^{-2}$) and 
($\mu_{Li}, n_{Li}(0)$) = (-11.15 K, 0.06 $\rm \AA^{-2})$.
These values  coincide with the coordinates of the respective
prewetting points shown on the curves for the chemical
potentials displayed in Fig. \ref{fig1}.

\section{Summary and outlook}

In this work we have described the spreading of $^4$He
droplets on alkali subtrates as they grow in size. One 
outcome of this investigation is that for wetted substrates,
the large size limit is not the bulk liquid
but, interestingly, the minimum stable film. In other words,
the 
deposited  cluster can grow towards the thermodynamic limit along the
directions permitted by the geometrical constraint, namely those
parallel to the planar substrate. The  transverse confinement just
fixes the maximum height of the sample  so as to yield the
prewetting coverage. 

Along this work we have shown density profiles of clusters 
on alkali adsorbers, and seen that for wetted 
substrates, droplets of $^4$He  atoms are present in the
  nonwetting, subspinodal regime 
-i.e., $\partial \mu/\partial n < 0$- and seem to present a contact 
angle in spite that, in the light of the results
  discussed in the previous section,  the asymptotic 
 limit of these large systems is the prewetting film, that is
  associated to vanishing contact angle. 
This apparent inconsistency stems from the impossibility to
define a  contact angle for nanoscopic droplets in which
the particle density displays a non-negligible surface width,
and is highly stratified near the substrate, which render the
contact angle an ill defined quantity. It also
illustrates that a determination of the contact angle by `visual
inspection' of the equidensity lines is fraught with danger and
may lead to either a crude estimate or even to a qualitatively
wrong result. As we have commented, Ancilotto {\it et al}\,
\cite{ancilotto98} presented  density  profiles of  $^4$He
droplets  on  Cs, at zero temperature, for the OT
density  functional.\cite{dalfovo95} 
In this work, emphasis lay on the determination of the contact angle
of macroscopic nonwetting samples of $^4$He on Cs, and a rather
complex procedure
was developed that permitted to compute this angle extrapolating 
the results for nanoscopic drops to the macroscopic ones  for
which a contact angle can be sensibly defined.
 We want to point out that since the FRDF here employed
differs only slightly from the OT -we have verified that no visible 
discrepancies show up in the structural properties of
clusters on Cs- applying the method of Ref. \onlinecite{ancilotto98} 
would yield in our case a similar contact angle for $^4$He on Cs
(about 36$^o$, cf. Table I in Ref. \onlinecite{szybiszp}).

An interesting extension of this work is the
investigation of the spreading of mixed $^3$He-$^4$He
clusters on planar substrates. Such a project is feasible within the 
FRDF formalism, and in fact
 we have recently shown\cite{mayol03} that one or few $^3$He atoms
 added to a large -yet nanoscopic-   
 $^4$He drop on Cs, localize on a onedimensional
ring around the contact line. This is a new feature of helium
mixtures, which may contribute  to  interpreting the very complex
phase diagram of such systems.\cite{ross96} A detailed systematics of
the structure and energetics of mixed clusters is presently in
progress  and will be reported elsewhere.

Finally, we would like to mention that the FRDF formalism may also be
used to shed light on the nucleation of wetting layers in the case of
a first-order wetting transition, very much as it has been successfully 
used to
address nucleation or cavitation in bulk liquid helium.\cite{bar02}
A major advantage of the formalism is that it may circumvent some of the
approximations made in other approaches. In particular, it avoids
the use of an experimentally unknown line tension,\cite{ind94} and
the use of a separation between surface and line contributions to
the free energy,\cite{blo95} which becomes doubtful when the radius of the
critical droplet or bubble is comparable to its surface width.

\acknowledgements

This work has been performed under grants BFM2002-01868  from
DGI, Spain, 2001SGR-00064 from Generalitat of Catalonia, 
EX-130 from University of Buenos Aires, and PICT2000-03-08450 
from ANPCYT, Argentina. E.S.H. has been also funded by M.E.C.D. (Spain)
on sabbatical leave.  We are grateful
to Francesco Ancilotto and to Milton Cole for useful discussions.

\begin{table}
\caption{
Energy 
${\cal E}_4$ 
(K) of one $^4$He atom on different alkali substrates.
}

\begin{center}
\begin{tabular}{|c|c|c|c|c|} 
    Li   &  Na  &  K   &  Rb  &  Cs  \\
\hline
  -10.70  &  -7.08 & -4.20 &  -3.72 &  -3.53  \\
\end{tabular}
\end{center}
\label{Tab1}
\end{table}

\begin{figure}
\caption[]{ 
Chemical potential $\mu$ (K) of $^4$He atoms in
 films upon Cs, Rb, K, Na and Li substrates  as functions of areal
 coverage $n$ (${\rm \AA^{-2}}$),
predicted by the FRDF employed in this work. Also shown is the
energy per particle $e$ (K) for Cs, Na, and Li, and the
energy per particle in bulk $^4$He $e_B$ as a straight line.
 }
\label{fig1}
\end{figure}

\begin{figure}
\caption[]{ 
Same as Fig. \ref{fig1} for the
surface grandpotential.
 }
\label{fig2}
\end{figure}

\begin{figure}
\caption[]{ 
Surface grandpotential as a function of inverse coverage.
The short horizontal line near $1/n=0$ represents the
value of the surface grandpotential in the limit of
infinite coverage.
 }
\label{fig3}
\end{figure}

\begin{figure}
\caption[]{ 
Contour plot of the density of a $^4$He$_{1000}$
droplet on Cs, K, Na and Li, from top to bottom.
The lowest equidensity lines correspond to
$\rho = 10^{-4} {\rm \AA}^{-3}$, and the highest
ones to
$ 2.6 \times 10^{-2} {\rm \AA}^{-3}$ (Cs and K), and
to
$ 3 \times 10^{-2} {\rm \AA}^{-3}$ (Na and Li).
 }
\label{fig4}
\end{figure}

\begin{figure}
\caption[]{ 
Density profiles $\rho(r, z_{min})$ (upper panel)
and $\rho(0, z)$ (lower panel) of a
 $^4$He$_{1000}$ cluster on different alkali substrates.
 }
\label{fig5}
\end{figure}

\begin{figure}
\caption[]{Chemical potential (left panel)
and grandpotential
per particle (right panel) of  $^4$He droplets on
alkali adsorbers as a function of $N$.
The energy per particle in bulk $^4$He $e_B$
is also shown in the left panel.
The lines have been drawn to guide the eye.
 }
\label{fig6}
\end{figure}

\begin{figure}
\caption[]{ 
Contour plots of the density of a $^4$He$_N$
 droplet on Na for $N=$ 100, 1000, 2000, and 3000.
The lowest equidensity lines correspond to
$\rho = 10^{-4} {\rm \AA}^{-3}$, and the highest
ones to  $3 \times 10^{-2} {\rm \AA}^{-3}$.
 }
\label{fig7}
\end{figure}

\begin{figure}
\caption[]{ 
Density profiles $\rho(r, z_{min})$ (upper panel)
and $\rho(0, z)$ (lower panel) of $^4$He$_N$ clusters on Na.
 }
\label{fig8}
\end{figure}

\begin{figure}
\caption[]{ 
Contour plots of the density of a $^4$He$_N$
 droplet on Li for $N=$ 100, 1000, 2000, and 3000.
The lowest equidensity lines correspond to
$\rho = 10^{-4} {\rm \AA}^{-3}$, and the highest
ones to
$ 2.5 \times 10^{-2} {\rm \AA}^{-3}$.
 }
\label{fig9}
\end{figure}

\begin{figure}
\caption[]{ 
Coverage $n(r = 0)$ for $^4$He$_N$ droplets
on the various alkali substrates as a function of atom number.
The lines have been drawn to guide the eye.
 }
\label{fig10}
\end{figure}

\end{document}